\documentstyle[aas2pp4]{article}
\input{epsf.sty} 

 \mathcode`*="002A

\newbox\grsign \setbox\grsign=\hbox{$>$} \newdimen\grdimen \grdimen=\ht\grsign
\newbox\simlessbox \newbox\simgreatbox \newbox\simpropbox
\setbox\simgreatbox=\hbox{\raise.5ex\hbox{$>$}\llap
     {\lower.5ex\hbox{$\sim$}}}\ht1=\grdimen\dp1=0pt
\setbox\simlessbox=\hbox{\raise.5ex\hbox{$<$}\llap
     {\lower.5ex\hbox{$\sim$}}}\ht2=\grdimen\dp2=0pt
\setbox\simpropbox=\hbox{\raise.5ex\hbox{$\propto$}\llap
     {\lower.5ex\hbox{$\sim$}}}\ht2=\grdimen\dp2=0pt

\received{1 April 1996}

\lefthead{Stern \& Svensson}
\righthead{Time Profiles of Gamma Ray Bursts}

\begin{document}

\title{Evidence for ``Chain Reaction'' in the\\
 Time Profiles of Gamma Ray Bursts}

\author{Boris E. Stern\altaffilmark{1,2} and 
Roland Svensson\altaffilmark{2}}\ 


\altaffiltext{1}{Institute for Nuclear Research, Russian Academy of Sciences
Moscow 117312, Russia, I: stern@al20.inr.troitsk.ru}
\altaffiltext{2}{Stockholm Observatory, S-133 36 Saltsj\"obaden, Sweden, 
I: svensson@astro.su.se}
 

\begin{abstract}
Although the time profiles of gamma ray bursts (GRBs)
show extremely diverse behavior, their average statistical properties such
as the average peak-aligned profile and the auto-correlation function
show simple stretched-exponential behavior. This could indicate that the diversity
of all bursts is just due to different random realizations 
of the same simple stochastic process where the process is 
scale invariant in time.
We illustrate how both the diversity 
of GRB time profiles and some important average statistical properties
can be reproduced in this way using a simple 
toy model for a stochastic {\it pulse avalanche}, which behaves as
a chain reaction in a near-critical regime.
We suggest that one possibility for the underlying physical 
process for generating  GRBs could be a ``chain detonation''
in which reconnecting magnetic turbulent features trigger each other.
\end{abstract}




%
  
\section{Introduction}
\label{sec:intro}

The temporal properties of GRBs in the hard X-ray to soft gamma ray 
band (25 keV - 1000 keV)
which are important for our analysis can roughly be summarized as follows:

 - The duration of bursts varies by five orders of magnitudes from a few 
milliseconds to a few hundred seconds and their time profiles
are extremely diverse in their complexity and morphology 
(e.g., \cite{fis93,fis95}).

 - All individual time profiles can probably be considered as a sum of
a number of pulses of a more or less standard
shape - a sharp rise and a slower exponential decay (for morphologically 
simple events this is evident, see, e.g., \cite{nor96}).
The width of these pulses differs by four orders of magnitude.

 - There are very complex events showing a wide range of time scales within
a single burst. Many short bursts or short features in 
long bursts look like rescaled
versions of much longer events (e.g., \cite{nor95}). This could indicate 
multifractality of the time profiles of GRBs (e.g., \cite{mer95,yua96}).

The procedure for obtaining the average peak-aligned time profile of GRBs 
was pioneered by Mitrofanov et al. (1994, 1995, 1996)
(also see Norris et al., 1994, 1996).
Stern (1996), studying the 
average peak-aligned time profile of GRBs
in the BATSE-2 catalog, found that the profile
has a simple ``stretched'' exponential shape,
$<F/F_p> = \exp[-(t/t_0)^{1/3}]$, where $t$ is the
time since the peak flux, $F_p$, of the 
event, and $t_0$ is a constant ranging from 0.3 sec for strong bursts 
to $\sim$ 1 sec for dim bursts. This dependence of $t_0$ on brightness
could be interpreted as cosmological time dilation (e.g., \cite{pac92,pir92}). 
Here we only consider the average time profile of all GRBs. 
In Figure 1a,  we present the average peak-aligned
time profile of GRBs, $<F/F_p>$, (thin-line histogram) using the 
larger statistics of the BATSE-3 catalog. 
Fitting $<F/F_p> = c\exp[-(t/t_0)^{\nu}]$ where $c$ is an additional parameter, 
and excluding the first 64 ms bin (likely to be contaminated by
Poisson noise) gives $\nu= 0.32 \pm 0.03$ at the 90 \% confidence
level.
The average time profile does not show any
statistically significant deviations from the stretched 
exponential for over almost 2.5 orders of magnitude
in fractional flux, $<F/F_p>$, and for
three orders of magnitude in time (from 
0.1 s to 100 s).
Such a simple average time profile is remarkable considering the diverse and 
chaotic behavior of the individual time profiles of GRBs.

The stretched exponential law is known to describe some 
characteristics of scale invariant processes associated with fractals, 
e.g., turbulence (\cite{jen92,chi94}).
The simple statistical behavior of GRBs could mean that 
the diversity of all bursts is just due to 
different random realizations 
of the same simple stochastic process where the process is 
scale invariant in time.
The stretched exponential time profile for GRBs should 
not be interpreted as representing some average relaxation process in GRBs, 
but rather to be a statistical effect
reflecting the distributions of pulse widths and time lags between 
pulses. 
In this paper, we illustrate how the variety
of GRB time profiles can be reproduced using a simple 
toy model for a {\it pulse avalanche}, which behaves as
a chain reaction in a near-critical regime.
We also obtain satisfactory agreement with a number of the
observed temporal statistical
properties of GRBs including the stretched exponential shape of the 
average time profile. 

\section{A Stochastic Pulse Avalanche Model}
\label{sec:model}

Previous attempts with, e.g., shot noise models to model 
temporal properties have focused on only
a few properties such as the power spectra of single bursts (\cite{bel92}), or
the average auto-correlation function (\cite{int96}).
Here, we attempt to interpret the recently discovered
stretched exponential as well as 
several other temporal properties of GRBs by suggesting 
the following hypothesis:

- All GRBs can be described as different random realizations of the
same simply organized stochastic process within narrow ranges of the 
parameters of the process.

- The stochastic process should be scale invariant in time.
 
- The stochastic process works near its critical regime. 
This would explain the large 
morphological diversity of GRBs.
 
We also suggest that one possible candidate for the underlying physical
process responsible for the stochasticity is magnetohydrodynamic
turbulence, where the standard pulses are due to magnetic reconnection events.

It is difficult, if not impossible, because of the peak-aligning
procedure to interpret the shape of the time profile using analytical 
methods. Instead we use numerical methods to develop 
a simple stochastic toy model that have similar temporal properties
as GRBs. 
 
The model we propose is a {\it pulse avalanche}, which is a linear Markov
process having the following properties: 

1.) The elementary event is a pulse with a time constant ($\approx$ pulse width)
$\tau$ and a standard shape being parameterized as a Gaussian rise, 
$A\exp[-(t-t_p)^2/\tau_r^2]$ for $t< t_p$, and an exponential decay, 
$A\exp[-(t-t_p)/\tau]$ for  $t > t_p$,
where $t_p$ is the time for the peak of the pulse.
Fitting of observed pulses gives $\tau_r/\tau \sim 0.3-0.5$ (Norris et al. 1996).
We use $\tau_r/\tau = 0.5$.
The pulse amplitude, $A$, is sampled from a uniform distribution,
$p_1(A)=1$, in the range [0, 1].
  
2.) In a pulse avalanche, each pulse acts as a parent pulse
giving rise to a number of baby pulses, $\mu_b$, 
sampled from a Poisson distribution, $p_2(\mu_b)= \mu^{-1}\exp(-\mu_b/\mu)$,
with the average number being $\mu$. The process is close to its critical
runaway regime when $\mu$ is of order unity.

3.) A baby pulse is assumed to be delayed a time, $\Delta t$,
with respect to the parent pulse. 
We parameterize the probability distribution for the Poisson  delay as
$p_3(\Delta t) = (\alpha \tau)^{-1} \exp(-{\Delta t}/{\alpha \tau})$,
where $\tau$ is the time constant of the baby pulse, and $\alpha$
is the delay parameter.
 
4.) How is the time constant of a baby pulse,
$\tau$, related to the time constant,  $\tau_1$, of the parent pulse?  
Studying individual time profiles we arrive at the intuitive conclusion that 
$\tau$ and $\tau_1$ are of the same order of magnitude and that
$\tau < \tau_1$ on average. This  allows the process to 
converge even if $\mu$ exceeds 1 as the pulse avalanche eventually
reaches an arbitrary short time scale, where a natural frequency cutoff should 
exist. We parameterize the probability distribution of $\log(\tau/\tau_1)$ as  
uniform, $p_4[\log(\tau/\tau_1)]=\mid \delta_2 -\delta_1 \mid^{-1}$ ,
in the range [$\delta_1$, $\delta_2$], where $\delta_1 <0, 
\delta_2 \geq 0$, and $\mid \delta_1 \mid > \mid \delta_2 \mid$.

5.) The process terminates when it converges due to subcritical values of 
the model parameters. 

6.) The start of the pulse avalanche must also be described.
We allow for the existence of 
a number, $\mu_s$, of spontaneous primary pulses,
sampled from a Poisson distribution, $p_5(\mu_s)= \mu_0^{-1}\exp(-\mu_s/\mu_0)$,
with the average number of spontaneous pulses per GRB being $\mu_0$. 
 
7.) We suggest that 
the probability distribution of the time constants, 
$\tau_0$, of spontaneous pulses is $p(\tau_0) \propto 1/ \tau_0$. This 
corresponds to flicker noise, i.e. a ``$1/f$'' spectrum, 
which is surprisingly wide spread in very
different classes of phenomena (e.g., \cite{pre78}).
Observations imply an upper cutoff, $\tau_{\rm max}$, for $\tau_0$. 
We then sample $\log \tau_0$ uniformly between $\tau_{\rm min}$ and 
$\tau_{\rm max}$, i.e. $p_6(\log \tau_0) = 
(\log \tau_{\rm max} - \log \tau_{\rm min} )^{-1}$,
where $\tau_{\rm min}$  should be smaller than the 
time resolution. Varying  $\tau_{\rm max}$ simply rescales
all average avalanche properties in time, in this sense,  $\tau_{\rm max}$ 
is a trivial parameter.

8.) The spontaneous primary pulses in a given GRB are all assumed to be delayed 
with different time intervals, $t$,
with respect to a common invisible trigger event.  
We parameterize the probability distribution for the Poisson  delay, $t$, of
a given spontaneous pulse as
$p_7(t) = (\alpha \tau_0)^{-1} \exp(-{t}/{\alpha \tau_0})$,
where $\alpha$
is the constant delay parameter used for all pulses 
(see property 3 above) and
$\tau_0$ is the time constant of the spontaneous pulse.
Each spontaneous
pulse gives rise to a pulse avalanche, and it is the overlap of $\mu_s$
pulse avalanches that form a GRB. 
(Alternatively, we could have chosen the spontaneous pulses to
appear  more or less uniformly over a wider time range that
characterizes the evolution time scale of the pulse generating object. 
The difference between the scenarios appears when the number of
spontaneous pulses exceeds unity.)
 
\section{Results of Simulations}
\label{sec:results}

The number of model parameters is seven. We found, however,
that it is comparatively easy to understand the effects of moving
around in parameter space.
First, we show how the model works for a certain set of parameters, that 
was chosen without any serious efforts of  optimization: $\mu=1.2$ (chosen by
visually examining when model bursts show sufficient complexity), 
$\alpha=4$ (gives the wanted shape of the average time profile),
$\delta_1 = -0.5$ and $\delta_2 =0$ (somewhat arbitrary), 
$\mu_0 = 1$ (arbitrary but reasonable),
$\tau_{\rm min} = 20$ ms (below the time resolution, 64 ms),
$\tau_{\rm max} = 26$ s
(by finding agreement with the experimental value, $t_0 = 0.82$ s, see Fig. 1a). 


Figure 1a tests the shape of the average peak-aligned 
time profile of the model. 
The  deviations over long time intervals 
between real and simulated profiles does not 
exceed 1.5 $\sigma$ and are typically within 
1$\sigma$ (where $\sigma = F_{\rm rms}/\sqrt{598}$
is the rms statistical error of the real average time profile 
for 598 useful BATSE-3 events and where the rms deviation 
$F_{\rm rms}$ is shown in Fig 1a). 
 This indicates that the two profiles
are statistically consistent (an exact
quantitative test for statistical consistency is very difficult due to
complicated correlations between different parts of the profile).  
An excellent stretched exponential shape is obtained for
$\alpha$ between 3 and 6 and the dependence on other 
parameters is weak. Outside this interval of $\alpha$,
the stretched exponential shape breaks at the level,
$<F/F_p> \sim 10^{-1.5}$, which would contradict the data.
Thus the long stretched exponential behavior is not an intrinsic
property of the model, but is achieved for a certain range of the 
delay parameter $\alpha$.
The observed time constant, $t_0 = 0.82$ s, of the average time profile
seems to be related to the geometric mean 0.72 s of $\tau_{\rm max}$
and $\tau_{\rm min}$

After the average time profile is fitted by varying only
two parameters, $\alpha$ and $\tau_{\rm max}$,
we automatically obtain agreement with observed data using a number of 
other characteristics:

 -- The root mean square deviations of individual profiles
(Fig. 1a). The agreement means that the model not only correctly reproduces 
the average time profile shape but also the fluctuations of individual profiles.
 
 -- The shape of the average peak-aligned distribution of time profiles to
the third power (Fig. 1b). 
  
-- The average auto-correlation function (ACF), (Fig. 1c). The 2 -- 10 s
excess of the real ACF over the simulated ACF
is statistically significant ($\sim 3\sigma$).
The deviation is moderate ($\sim$ 13\%), however, and thus the test is reasonably
successful.  

-- The duration distribution (Fig. 1d). Here we claim only approximate agreement
as discussed in the figure caption.

In Figure 2,  four observed time profiles are compared with
simulated time profiles of similar complexity and morphology. Counterparts
for each real event were sampled from 300 simulated events using a {\it single} 
set of model
parameters fixed to the values given above. The exception is the 
counterpart for the most erratic event in the BATSE-3 sample (Fig. 2d) 
where one parameter was changed slightly: 
the criticality was increased by setting $\delta_2$= 0.2 instead 
of 0.

A visual examination of burst profiles shows that the model is reasonably
successful.
For a more quantitative comparison, we selected the 325 brightest BATSE-3
bursts and simulated the same number of events using the set of parameter
values given above. In order to eliminate brightness selection effects, we
wanted the amplitude and background distributions of real and simulated
events to be the same.
Each simulated event, therefore, had a peak and a background count rate of 
the $i$-th real event, where $i$ is a  random number, $0 < i < 325$. 
Furthermore, we imposed Poisson noise at the 64 ms time resolution. 
Then we made a simple visual morphological classification of both real
and simulated bursts using the same criteria.
We found 104 real and 91 simulated {\it single peaked} events, 43 and 44
{\it double peaked} events, 84 and 81 {\it moderately complex} events,
94 and 109 {\it erratic} events. 
Repeating the same test for $\mu_0=0.5$ instead of $\mu_0=1$ gave
112, 46, 79 and 88 simulated events in each class, respectively. 
Keeping in mind that visual classification is very subjective, especially between
moderately complex and erratic events, we note that both results are 
statistically consistent with the distribution of real bursts.
We also note that the actual number of simulated pulses in a GRB 
often exceeds the visually estimated number.
 
We still have a number of parameters to use for fine-tuning the model,
but obtaining the best fit is beyond the scope of this letter. We just 
present support for the hypothesis formulated above by using
a simple model to demonstrate that
a number of temporal properties of GRBs have a natural interpretation in
this approach. We cannot and do not claim that we have found the only possible
model that satisfy the data, but it is probably the simplest possible one. 
Such basic features as a $1/f$ spectrum,
time lag between pulses as described by property 3) in \S~\ref{sec:model}, and 
time scaling invariance should 
probably be present in any model.

The model we present is a version of a chain reaction in a near-critical
(slightly subcritical)
regime which then naturally provides large fluctuations. 
One other feature -- the $1/f$ spectrum of spontaneous pulses -- could also be
associated with the near-criticality of the system. 
This near-criticality could be related to the concept of 
self-organized criticality that was introduced by  
Bak, Tang, \& Wiesenfeld (1987) 
in order to explain the widespread occurrence of $1/f$ noise.

\section{Discussion}
\label{sec:discussion}

We now discuss the model in terms of the possible
underlying physical process. We suggest  
reconnecting magnetic turbulence as a possible underlying mechanism.
Magnetic reconnection can generate 
abrupt releases of huge amounts of energy into hard X-rays or soft gamma rays.
This is the case for solar flares and probably also for AGNs
(\cite{gal79,haa94}).   
Turbulence is a phenomenon showing spatial and temporal scaling invariance, 
with the larger scales cascading to smaller scales just as required in our model.
 
Consider the following illustrative example for a possible scenario for GRBs 
which incorporates the pulse avalanche. 
The original trigger event could be the coalescence of two neutron stars 
(e.g., \cite{nar92})
or a catastrophic energy release by a neutron star (e.g., \cite{uso92}).
The trigger event itself can be invisible in the BATSE range
on the dynamical time scale of the event (\cite{mes93}).
Almost all released energy goes into the kinetic energy
of the expanding fireball (e.g., \cite{cav78,pac90}). 
Then, at some stage, instabilities develop
in the fireball due to interactions with the interstellar medium
(e.g., \cite{ree92,mes93})
or due to collisions of blast waves (e.g., \cite{ree94}). 
Instabilities could lead to
the conversion of fireball energy into turbulent magnetic fields
and eventually to dissipation through magnetic reconnection.

These reconnection events can proceed as a chain 
detonation of turbulent features
 -- the reconnection of one feature destabilizes other features 
each of which will reconnect after some time delay.

The active turbulent phase can last for more than one hundred seconds, but if
the probability of spontaneous reconnection is not large ($\mu_0 \sim 1$),
we may see nothing (zero primary pulses), or just one single pulse 
(the chain is interrupted after the first pulse), 
or a developed complex avalanche -- it all depends on chance. 
The rest of the fireball energy (probably the major part) dissipates
in other directions or 
on longer time scales and in other energy ranges and is not yet detected.

The scenario described above is just 
an illustration and for this reason we did not check whether it satisfies 
necessary temporal and energy requirements. It is, however, probable that the
concept of a chain reaction can be built into a number of different scenarios.

There exist more than one hundred models of GRBs (\cite{nem94}). 
We are not suggesting a 
new model but rather a new approach to the interpretation of the temporal 
properties of GRBs, an approach that seems very fruitful and 
already extends our understanding of time profiles of GRBs.
 
\acknowledgments
 
We thank Andrej Doroshkevich, Ber\-nard Jones, Juri Poutanen,
and Martin Rees for useful dicussions. We also thank Juri Poutanen for
helpful assistance. We acknowledge support 
from the Swedish Natural Science Research Council and 
from a Nordita Nordic Project grant.



\clearpage

%
%

\clearpage
\onecolumn

\begin{figure} 
\leavevmode
\epsscale{.6}
\epsfysize=12cm  \epsfbox{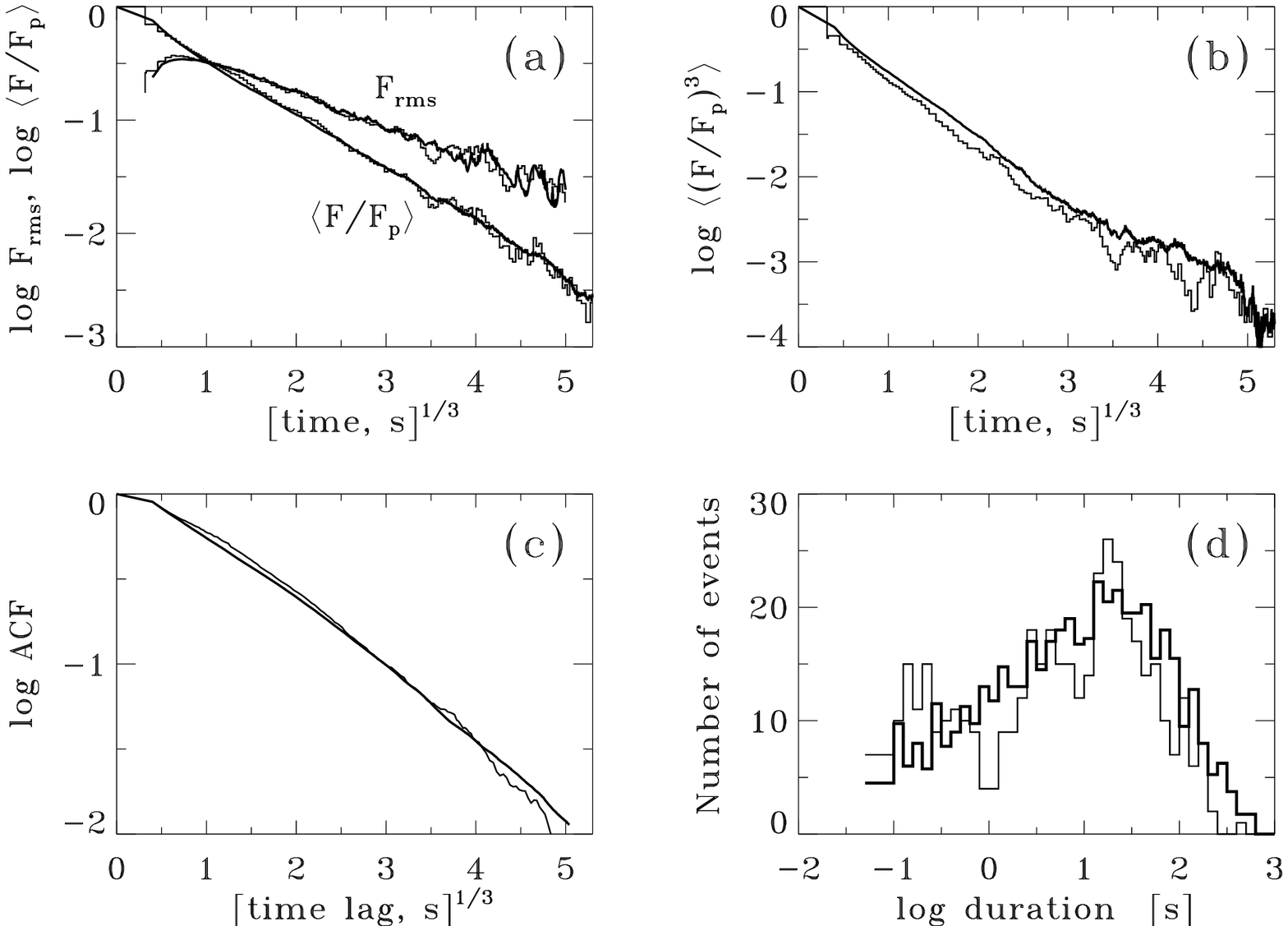} 
\caption{
Comparison between observed and simulated average temporal functions of
time profiles of GRBs.
(a) Thin-line histogram labelled $<F/F_p>$ shows the average 
peak-aligned post-peak 
time profile for the 598 useful BATSE-3 events 
as the fractional flux,
$<F/F_p>$ vs $t^{1/3}$, where $F_p$ is the peak flux and
$t$ = 0 -- 150 s is the time since the strongest peak.
Further details are given by Stern (1996).
Thick curve labelled $<F/F_p>$ is the average peak-aligned time profile
for 5000 simulated time profiles. The set of curves labelled
$F_{\rm rms}$ are the
rms deviations of individual peak-aligned time profiles,
$F_{\rm rms} \equiv (<(F/F_p)^2>-<F/F_p>^2 )^{1/2}$,
for
both real (thin line histogram, 598 events) and
simulated (thick curve, 5000 events) time profiles. 
(b) Third moment test, i.e. comparing the 
peak-aligned $<(F/F_p)^3>$.
The only significant difference between the curves is in the relative
height of the first bin, representing the peak of the individual time profiles. 
This is probably
due to the effect of Poisson noise which is enhanced here as compared to
the first moment profile in (a).
(c) Auto-correlation function (ACF) test. The definition of
the auto-correlation function for time profiles with Poisson noise are
given by
Link, Epstein, \& Priedhorsky (1993). 
The average ACF is approximately the same
for both real (thin curve, 598 events) and simulated (thick curve,
2000 events) time profiles, except for an excess in the real ACF for time lags
2-10 s (also visible for the average time profile in (a)). 
The excess
may be due to a larger correlation between pulses in real events.
(d) Duration distribution test.
The number of events of a given duration vs the duration time
for both real events (thin line, 432 sufficiently bright events) 
and simulated events (thick line,
2000 events with the distribution reduced by a factor four).
The durations of both real and simulated events were measured at a
level of 20 \%  of the peak amplitude.  Real events
were filtered to reduce the Poisson noise.
For the chosen set of parameters (see text),
the positions of the main peak of the simulated and 
observed  duration distributions are in remarkable agreement,
while the possible
double-humped shape  of the observed distribution which
has been interpreted as a bimodality of GRBs 
(Kouveliotou et al. 1993) 
is not reproduced.
A Kolmogorov-Smirnov test shows that these two distributions
are drawn from the same parent distribution with a probability of 0.007.
The bimodality, if it exists, can
be implemented into the model by having the primary
pulse width  spectrum deviate from $1/f$ flicker noise.
\label{fig1}
}
\end{figure}

\begin{figure}
\leavevmode
\epsfysize=12cm  \epsfbox{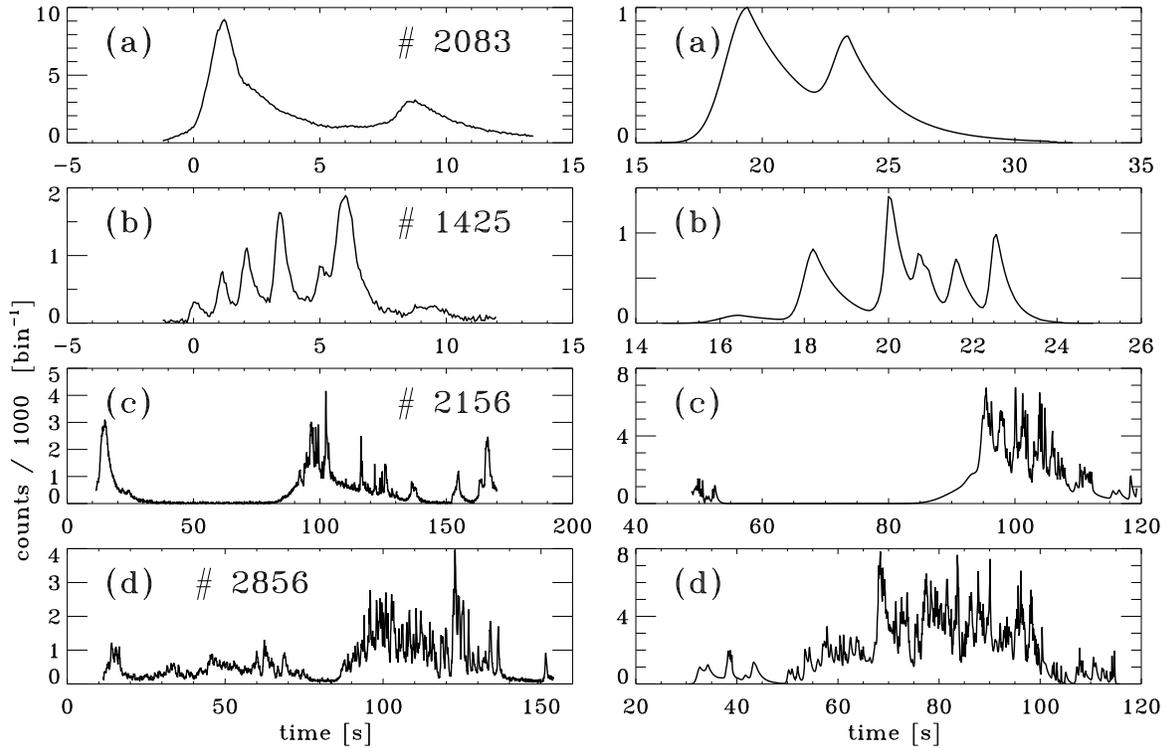}
\epsscale{.6}
\caption{
Observed time profiles of bright GRBs (left panels,
count rate vs time after BATSE trigger, trigger \# given in panels) and
simulated pulse avalanches (right, flux over the maximum amplitude
for a single pulse vs time after invisible trigger).
The four pairs of time profiles are selected to represent
some of the morphological classes discussed in the literature
(Fishman 1993, Fishman \& Meegan 1995):
(a) smooth profile, few peaks,
(b) smooth profile, blend of multiple peaks, 
the simulated event consists of 9 pulses, the
real event probably consists of about 10 - 15 pulses,
(c) complex profile, well separated episodes of emission, 
the simulated event consists of 395 pulses,
(d) very erratic, chaotic, and spiky bursts, the simulated one 
consists of 990 pulses,
the real one is probably the sum of more than 1000 pulses.
The parameters for the three simulations (a)-(c) are the same (see text),
For case (d), the criticality is slightly
increased by setting $\delta_2$ = 0.2 instead of 0.
Note that in complex simulated events (as well as in complex real events),
the apparent amplitude of spikes is higher in
dense pulse bunches as compared to single pulses despite
the fact that all simulated pulses have an amplitude sampled
uniformly between 0 and 1.
The reason is that the high spikes are the unresolved sum of many
correlated narrow pulses. The same is probably also the case for real events.
The simulations also reproduce observed GRB features such
as ``flat-top'' pulses and sharply terminated events, with
the latter being due to avalanche chains converging to very short
time scales.
\label{fig2}
}
\end{figure}

%

\end{document}